\documentclass[final]{aipproc}
\usepackage{graphicx}
\usepackage{epsf}

\usepackage{times}

\layoutstyle{8x11single}

\usepackage{bm}
\newcommand{\x}{\ensuremath{\bm{x}}}

\begin{document}
\title{Modern Statistical Methods for GLAST Event Analysis}

\author{Robin D. Morris}{
  address={USRA-RIACS, 444, Castro St, Suite 320, Mountain View, CA 94041}
}
\author{Johann Cohen-Tanugi}{
  address={SLAC/KIPAC, Stanford University, Stanford, CA}
}

\begin{abstract}
We describe a statistical
reconstruction methodology for the GLAST LAT.  The methodology
incorporates in detail the statistics of the interactions of photons
and charged particles with the tungsten layers in the LAT, and uses
the scattering distributions to compute the full probability
distribution  over the energy
and direction of the incident photons. It uses model selection methods
to estimate the probabilities of the possible geometrical
configurations of the particles produced in the detector, and
numerical marginalization over the energy loss and scattering angles
at each layer.  Preliminary results show that it can improve on the
tracker-only energy estimates for muons and electrons incident on the
LAT.
\end{abstract}

\maketitle

\section{Introduction}

The Large Area Telescope (LAT) \cite{LAT} is the primary instrument on GLAST, and
so it is of utmost importance to extract as much information as
possible from the response of the LAT to incident photons and
particles.  While the quantities of primary interest for each event are
few (namely the azimuth, elevation and energy of the incident
photon/particle), the rich physics of the interactions of the
particles/photons with the LAT makes a principled reconstruction
algorithm complex.  Whilst the objects of primary interest are
photons, the interaction of a photon with the LAT is essentially that of an
electron-positron pair. Therefore we concentrate on the basic
building blocks of event reconstruction, the analysis of the
interaction of charged particles with the LAT.

\begin{figure}
  \hspace*{0.05\textwidth}
  \includegraphics[width=0.2\textwidth]{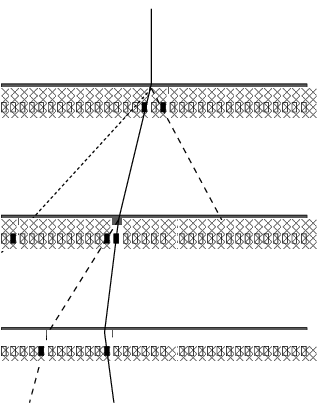}
  \hspace*{0.1\textwidth}
  \includegraphics[width=0.6\textwidth]{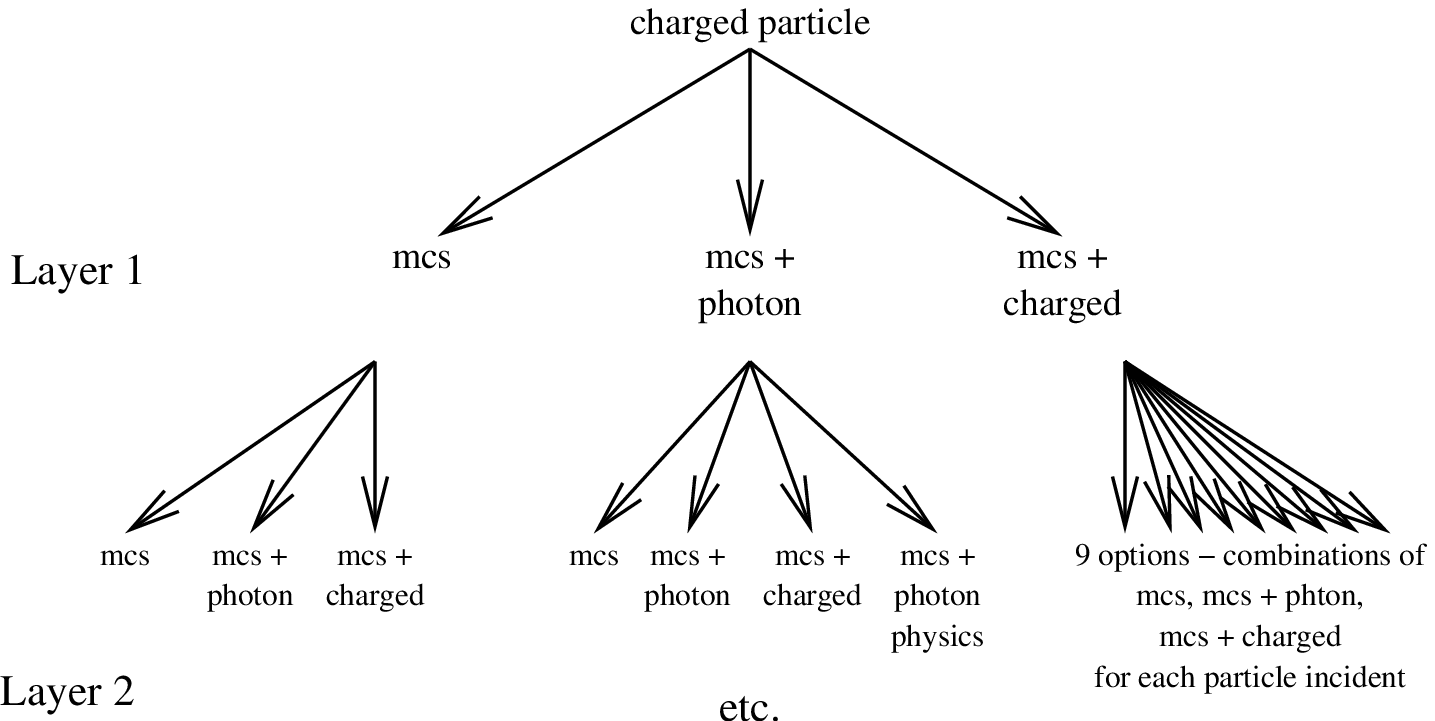}
  \hspace*{0.05\textwidth}
\caption{LEFT: A schematic of a charged particle interacting with the
  LAT. The solid line is the incident charged particle.  Long-dashed
  lines indicate secondary charged particles.  The short-dashed line
  is a secondary photon.  The solid blobs indicate microstrips that
  fired.  
  RIGHT: The tree of possible event structures}
\label{fig:tree}
\end{figure}

Figure \ref{fig:tree} (left) shows the schematic of an interaction between a
charged particle and the LAT.  Visible in the figure are 1) multiple
Coulomb scattering in the tungsten foils; 2) the production of
secondary photons; and 3) the production of secondary charged
particles.  The GEANT4 toolkit \cite{geant4} is designed to simulate these physics
processes in the {\em forward} direction.  The task in event
reconstruction is the {\em inverse} problem -- estimating, from the
data of the microstrip responses, which physics processes actually
occurred in a particular event.  The result is an estimate not only for
the original particle and its properties, but also of all secondary
particles and photons.  To accurately estimate the primary particle, it
is necessary to estimate accurately all secondaries.

Figure \ref{fig:tree} (right) shows the tree of hypotheses for the physical
processes at the first two layers of interaction of a charged particle
with the LAT. The final leaves describe the {\em structure} of the
hypothesized event reconstructions.
 Clearly as we descend the layers the number of branches
in this tree explodes.   But only a small number of their leaves will
be consistent with the instrument data - with the pattern of
microstrips that fired - and, further, this consideration can be used to exclude
branches  as the tree is descended, further limiting
the number of leaves that must be considered.  A full event
reconstruction consists of two stages:
\begin{enumerate}
\item The enumeration of the possible event structures consistent with
  the microstrip data.
\item The computation of the parameters of each event structure and
  their relative probabilities.
\end{enumerate}
Computation of the relative probabilities allows the final event
reconstruction to be a weighted average of event structures, weighted
according to the probability that the microstrip data actually came
from that structure.

\section{Methodology}

We concentrate first on the simplest type of event.  If the only
physical process that actually occurred was multiple Coulomb
scattering, then there is only a single (x-y) pair of microstrips at
each layer (excluding noisy strips).  This is the case for muons of
moderate energy (up to a few hundred MeV), where the probability of
producing secondary electrons or photons is extremely small, and hence
can be neglected.  We parameterize the trajectory of the particle by
1) its origin (a point outside the LAT); 2) the position at which it
traverses each conversion layer; and 3) its endpoint (also outside the LAT);
and from these we derive the incident directions $(\theta,\phi)$ and
the scattering angles, $\theta_i$, at each layer.  Finally, we add the
incident energy, $E$, and the energy deposited in each conversion
layer $\delta E_i$.  Denoting by $s_i$ the microstrips at each layer,
we can write
\begin{equation}
p(\theta,\phi,E,\theta_1,\delta E_1, \ldots \theta_n,\delta E_n | s_1
\ldots s_n) 
\propto
p(s_1\ldots s_n | \theta,\phi,E,\theta_1,\delta E_1, \ldots
\theta_n,\delta E_n)
p(\theta,\phi,E,\theta_1,\delta E_1, \ldots \theta_n,\delta E_n)
\label{eq:bayes}
\end{equation}
The first term on the right hand side is the likelihood.  It takes one
of two values -- one if the trajectory described by
$\theta,\phi,\theta_1,\ldots \theta_n$ intersects all the microstrips
that fired, and zero otherwise.  It serves to limit the region of the
state space that is of interest.  The second term contains all the
physics of the interactions of the particle with the LAT.  We use
conditional independence and decompose it as follows.
\begin{eqnarray}
p(\theta,\phi,E) & & \mbox{Priors on azimuth, elevation and energy.}
\nonumber \\
\times p(\theta_1,\delta E_1 | E ) & & \mbox{Distribution of scattering angle
  and energy loss for a particle of energy $E$.} \nonumber \\
\times p(\theta_2,\delta E_2 | E, \delta E_1) & & \mbox{Same for the particle
  at layer 2, which has energy $E-\delta E_1$.} \nonumber \\
\times \ldots & & \nonumber \\
\times p(\theta_n,\delta E_n | E, \delta E_1, \ldots, \delta E_n) & & \mbox{At
layer $n$ the particle has energy $E-\delta E_1 - \ldots - \delta E_{n-1}$.}
\end{eqnarray}
These scattering distributions are the known distributions for
particles of a specified energy incident on a LAT foil \cite{poptm}.
The 
non-Gaussian 
tails of the scattering angle distributions were
modeled by a second Gaussian component.  For muons, the energy loss
was parameterized as a Landau distribution, with the distribution's
parameters being functions of energy.

The parameters of primary interest, however, are the azimuth,
elevation and energy, and so the distribution of primary interest is
$p(\theta,\phi,E)$.  This is obtained from (\ref{eq:bayes}) by
marginalization.  This is performed numerically using Markov chain
Monte Carlo (MCMC).  Originally developed in physics \cite{metropolis}, it
has been extensively developed in statistics in the past 20 years, and
is now a standard tool for use in the analysis of complex,
high-dimensional probability distributions \cite{mcmc}.  It works by simulating a
Markov chain whose equilibrium distribution is constructed to be the
distribution of interest (in this case, the distribution in
(\ref{eq:bayes})).  Averages over the distribution can then be made by
forming averages over the states of the simulated Markov chain.  For
example, the mean energy is estimated by forming the mean of the
energy variables over a length of the simulated chain, while ignoring
all the other variables.  Collecting all the variables into \x,  and
initializing $\x\leftarrow\x_0$
the 
MCMC algorithm is iteration of
\begin{enumerate}
\item propose a change, $\x\leftarrow\x'$ with some proposal
  distribution $\pi(\x';\x)$
\item accept the change with $p_a =
  \frac{p(\x')\pi(\x;\x')}{p(\x)\pi(\x';\x)}$, and set
  $\x\leftarrow\x'$, else retain \x
\end{enumerate}
The proposal distribution at each stage may be chosen to only change
some of the elements of \x.  For this work, we use a cycle of
proposals that successively proposes changes to
$\theta,\phi,E,\theta_1,\delta E_1,\ldots,\theta_n,\delta E_n$.

For electrons incident on the LAT foils, as well as multiple Coulomb
scattering, there is appreciable probability of producing a secondary
photon, and a small probability of producing a secondary electron. (At
100MeV these probabilities are $\simeq 0.25$ and $\simeq 0.01$
respectively.)  We restrict the discussion here to events which
contain at most secondary photons.  Typically, for electrons of a few
hundred MeV the secondary photons are not detected.  They carry energy
away from the electron which is ``lost'' to the tracker.

In the forward direction this is modeled by a mixture distribution.
With probability $p_{\mbox{ns}}(E)$ no secondary is
produced, and the energy loss follows a Landau distribution.  With
probability $p_{\mbox{s}}(E)$, a photon is produced and the
energy loss has two components, a Landau distributed component from
multiple Coulomb scattering, plus a component distributed as $1/E$
representing the energy carried away by the photon.

The samples generated by the MCMC algorithm represent the distribution
over the trajectories' parameters.  To compute the probability for an
event structure it is necessary to compute the normalizing factor that
was omitted from equation (\ref{eq:bayes}).
This can be done by using the MCMC output to construct an importance
sampling distribution, and using samples from that distribution to
compute the normalizing factor.  This will be discussed elsewhere.

\section{Results}

\begin{figure}
  \vspace*{0.05\textwidth}
  \includegraphics[width=0.4\textwidth]{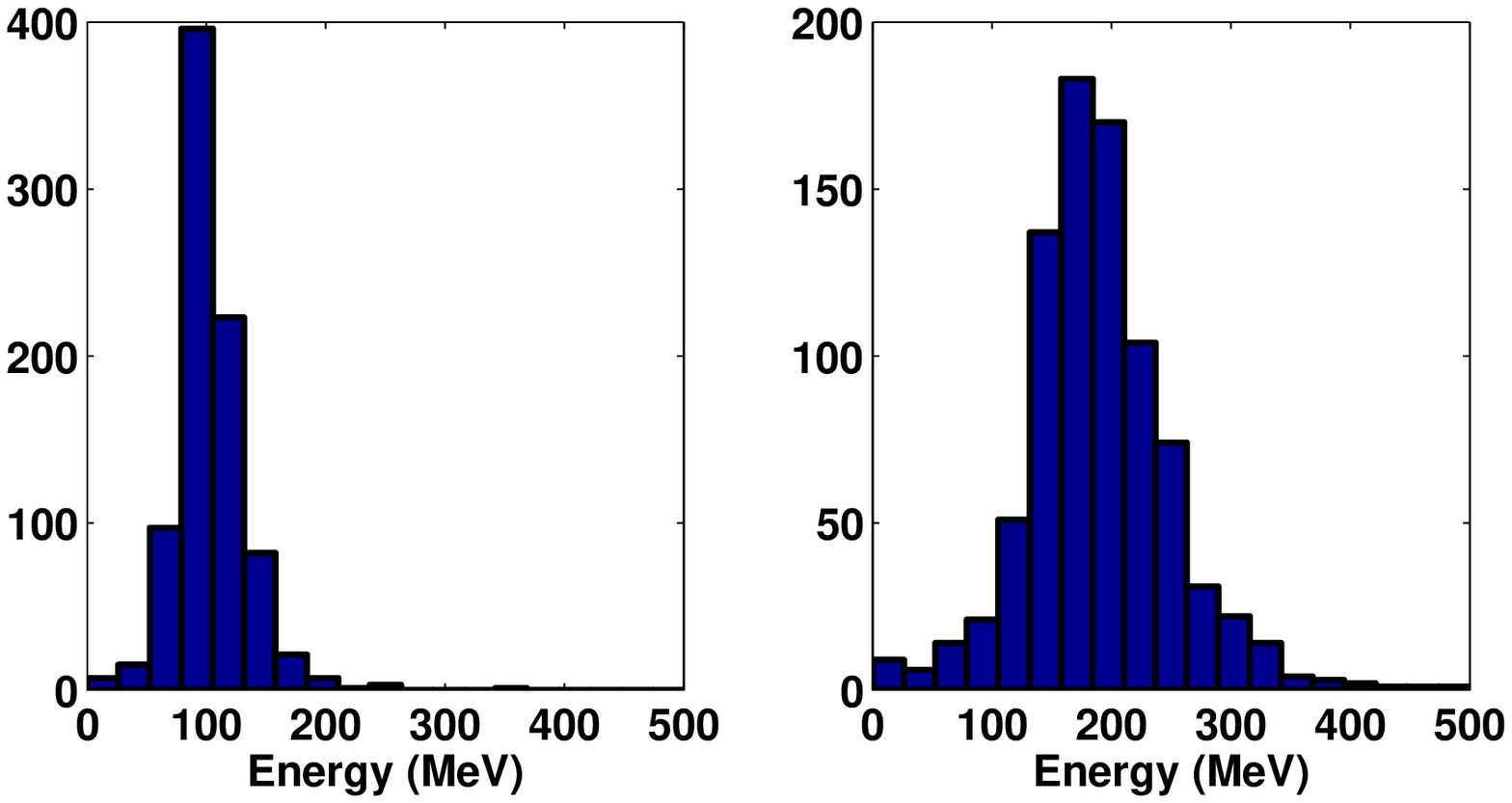}
  \vspace*{0.1\textwidth}
  \includegraphics[width=0.4\textwidth]{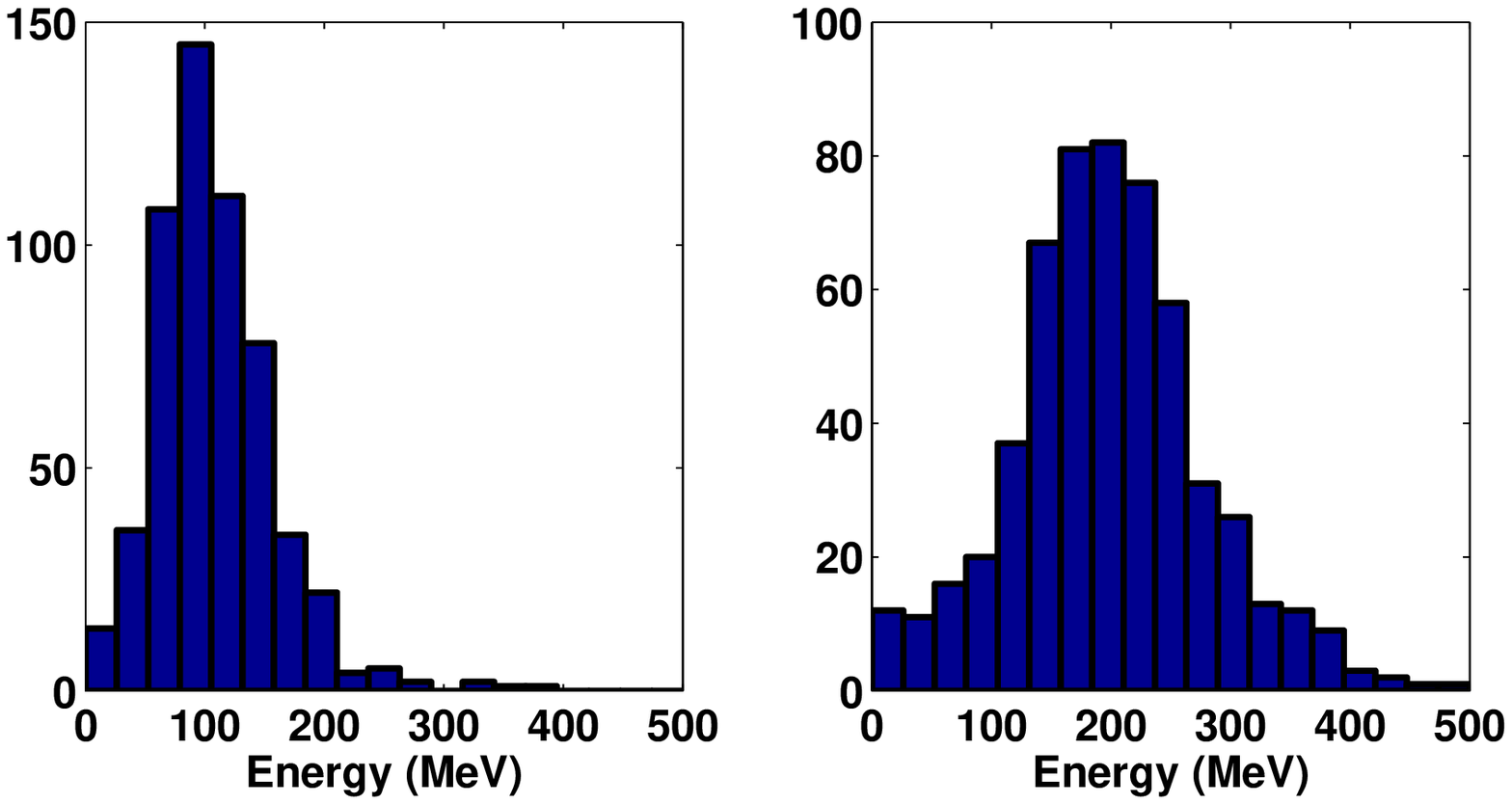}
  \vspace*{0.05\textwidth}
  \caption{Energy estimates for muons (2 left panes) and electrons (2
    right panes)}
  \label{fig:results}
\end{figure}

One thousand events were simulated \cite{latsim} for each of four
sources -- muons  and electrons of
100 and 200 MeV.  The incidence
directions were chosen randomly within a 45 degree cone.  Figure
\ref{fig:results} shows the energy estimates of the reconstruction. 
We do not show direction estimates for reasons of space and also
because, for charged particles, the accuracy of the direction estimate
is determined almost entirely by multiple Coulomb scattering in the
top foil.
For
  electron events, those events where two charged tracks were detected
  by the reconstruction algorithm were not analyzed.  In all four
cases the estimates are unbiased; the histograms are centered
accurately on either 100MeV or 200MeV.  The histograms for electrons
show more dispersion than those for muons, due to the effect of
energy being transferred into photons which are not detected.  Note
that these estimates were made using only the first 12 regular GLAST
layers of the tracker and did not use any information from the calorimeter. 
In an upcoming paper we will present detailed results
and comparisons with current methodology.

\begin{theacknowledgments}
Our analysis methodology is built upon the software provided by the
LAT Science Analysis Software Team.
RDM is supported by a grant from the NASA AISR program.
\end{theacknowledgments}

\end{document}